\documentclass[prl,aps,twocolumn,superscriptaddress,longbibliography]{revtex4-1} 
\usepackage[normalem]{ulem}
\usepackage{graphicx} 
\usepackage[usenames]{color}
\usepackage{amsmath,amssymb}
\usepackage{wasysym}
\usepackage{gensymb}
\usepackage{bbold}
\usepackage{natbib}
\usepackage[%
  colorlinks=true,
  urlcolor=blue,
  linkcolor=blue,
  citecolor=blue
]{hyperref}

\usepackage[usenames,dvipsnames]{xcolor}
\usepackage{lipsum}
\usepackage{color}

\begin{document} 
\title{Forced imbibition in porous media: a fourfold scenario}
\author{C\'eleste Odier}
\affiliation{Univ Lyon, Ens de Lyon, Univ Claude Bernard, CNRS, Laboratoire de Physique, F-69342 Lyon, France}
\affiliation{Total SA. P\^ole d'Etudes et Recherche de Lacq, BP 47-64170 Lacq, France}
\author{Bertrand Levach\'e}
\affiliation{Total SA. P\^ole d'Etudes et Recherche de Lacq, BP 47-64170 Lacq, France}
\affiliation{Laboratoire Physico-Chimie des Interfaces Complexes, Total-ESPCI Paris-CNRS-UPMC, BP 47-64170 Lacq, France}
\author{Enric Santanach-Carreras}
\affiliation{Total SA. P\^ole d'Etudes et Recherche de Lacq, BP 47-64170 Lacq, France}
\affiliation{Laboratoire Physico-Chimie des Interfaces Complexes, Total-ESPCI Paris-CNRS-UPMC, BP 47-64170 Lacq, France}
\author{Denis Bartolo}
\affiliation{Univ Lyon, Ens de Lyon, Univ Claude Bernard, CNRS, Laboratoire de Physique, F-69342 Lyon, France} 
\date{\today} 

\begin{abstract}
We establish a comprehensive description of the patterns formed when a wetting liquid  displaces a viscous fluid confined in a porous medium. Building on model microfluidic experiments, we evidence four imbibition scenarios all yielding different large-scale morphologies. Combining high-resolution imaging and confocal microscopy, we show that they originate from two liquid-entrainment transitions and  a Rayleigh-Plateau instability at the pore scale. Finally, we demonstrate and explain the long-time coarsening of the resulting patterns.

%

\end{abstract} 
\maketitle

About a hundred years ago field engineers noticed  that water  does not homogeneously displace  oil confined in porous media~\cite{Buckley42}. After a century of intense research,  our understanding of driven liquid interfaces remains surprisingly unbalanced.
The case of drainage where the defending fluid wets the confining solid,  is fairly well understood, and so is the spontaneous imbibition of a wetting liquid through an empty porous structure, such as the rise of a coffee drop in a sugar cube~\cite{Lenormand1897,Lenormand90,Sandnes2011,Sahimi93,Alava2004}. In stark contrast, much less attention has been paid to the situation where imbibition is used to mobilize a more viscous liquid~\cite{Chaikin86,Stokes87,Lenormand90}.  For more than twenty years, and despite its relevance to a number of industrial processes, our knowledge about forced imbibition  has been mostly limited to a seminal phase diagram constructed by Lenormand from heuristic arguments ~\cite{Lenormand90}. However,  recent microfluidic experiments~\cite{Levache2014,Wexler1,Juanes1,Juanes2}, and simplified 2D simulations~\cite{Segre2015} have  revisited this framework, and established the 
dramatic impact of wetting properties on  liquid-displacement patterns.
Nonetheless,  we still lack a comprehensive picture of forced-imbibition patterns, and of the underlying interfacial dynamics at the pore scale. 

In this letter, we contribute to resolving this situation. 
Building on model microfluidic experiments, we  evidence four classes of imbibition regimes yielding non-monotonic variations of the recovery rate with the driving strength.   We show that these  patterns stem from four different microscopic  dynamics separated by two film-entrainment transitions and one interfacial instability.  Finally, we demonstrate and explain the long-time coarsening of two classes of imbibition patterns.

\begin{figure*}
\begin{center}
  \includegraphics[width=\textwidth]{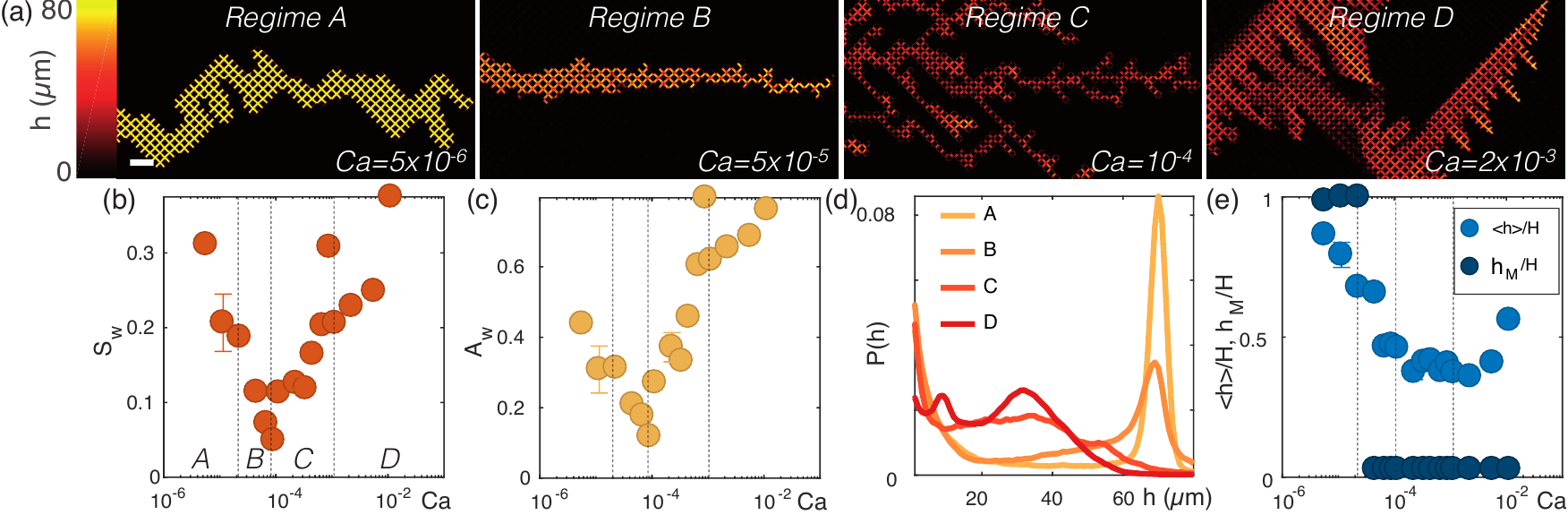}
  \caption{\label{fourregimes}
 Imbibition patterns.
  (a): Four imbibition patterns corresponding to the four regimes discussed in the main text. An aqueous solution is driven though  a periodic lattice of channels  filled with silicon oil (M=560). The color indicates the local thickness of the water films. Scale bar: $1\,\rm mm$. (b) Variations of the water saturation with $Ca$.   (c): Variations of the wetted area fraction with $Ca$. (d) Probability distribution of the local water-film thickness for the four capillary numbers corresponding to the experiments shown in (a). (e) Light blue circles: average thickness of the water films normalized by the average height of the microchannels $H$. Dark blue: most probable film thickness, $h_{\rm M}$ normalized by $H$.  Error bars: 1 $\sigma$ (four different experiments). The dashed vertical lines separate the four imbibition regimes. 
  }
\end{center}
\end{figure*}

Our experiment consists in injecting an aqueous liquid in a hydrophilic porous medium filled with  viscous  oil. Details about the  materials and methods are provided in a Supplementary Document~\cite{supp}. In brief,  
we use a simplified 2D geometry consisting of a square lattice of posts forming  interconnected channels of dimensions $200\,\rm \mu m\times80\,\mu m\times 70\,\mu m,$ see in Figs.~\ref{fourregimes}a. 
The patterns are imprinted on   microfluidic stickers   
bonded to quartz slides, and  made hydrophilic by in situ exposure to  deep-UV light~\cite{Bartolo2008,Levache2012}.
 We first fill the channels with a mixture of  silicon oils, which we then displace by injecting  an aqueous solution of dye and SDS ($1\,\rm wt\%$). The surface  tension between the two liquids does not depend on the oil molecular weight: $\gamma=13\pm 2\,\rm mN/m$,  and we find  advancing contact angles $\theta_{\rm A}=30\pm2^\circ$, and  $\theta_{\rm A}=20\pm2^\circ$, respectively  on the  treated quartz and sticker surface.
The water solution is injected at constant flow rate $Q$, and the large-scale morphology of the imbibition patterns is observed with a 14-bit, CCD-camera  with a spatial resolution of $11\,\rm \mu m/pixel$. We gain information about the interfacial morphology in the third dimension by locally converting the transmitted-light intensity  into the  $z$-averaged thickness of the water patterns with a relative precision of $1\,\rm \mu m$~\cite{Levache2014}. In addition, 3D reconstructions of the liquid interfaces at the pore scale are achieved in separate experiments  using  confocal microscopy. 
%

Let us first discuss the impact of flow rate on the pattern morphologies, while keeping constant both the wetting contact angles and the viscosity ratio, here $M\equiv\eta_{\rm oil}/\eta_{\rm water}=560$. Without any a priori knowledge of the microscopic dynamics, a  natural dimensionless control parameter is given by  the  capillary number ${Ca}=\eta_{\rm water} Q/(\gamma S)$, where  $S$ is the  cross-section area of the microchannels separating the posts. 
Increasing $Ca$, we observe four different  patterns exemplified in Fig.~\ref{fourregimes}a: 

\noindent{\em -- Regime A:}  When $Ca<2.3 \times 10^{-5}$, the patterns are typical of capillary imbibition~\cite{Lenormand90}. A single macroscopic finger proceeds at  constant speed, and its faceted edges reflect the square geometry of the underlying periodic lattice~\cite{Courbin2007,Cubaud2001}. 

\noindent{\em -- Regime B:} Increasing the flow rate, $2.3 \times 10^{-5}<Ca<9.1 \times 10^{-5}$, we still observe the growth of a single yet narrower finger. However, the dynamics of the water-oil interface does not  reduce to the motion of a stable meniscus. Instead, the spatial fluctuations of the transmitted light indicate that thin water films form and propagate throughout the lattice. 

\noindent{\em -- Regime C:} Further increasing the flow rate, $9.1 \times 10^{-5}<Ca<1.1 \times 10^{-3}$,  the patterns evolve into even thinner films  forming river networks with multiple narrow branches and reconnections. We emphasize that the existence of closed loops distinguishes this geometry from typical viscous fingering  motifs, and from all  instances of Laplacian growth patterns~\cite{Sahimi93}. 

\noindent{\em -- Regime D:} Finally when $Ca>1.1 \times 10^{-3}$,  water films self-organize into non-intersecting branched patterns. They are reminiscent of dendritic growth and drainage dynamics~\cite{Chen85,Sander85,Garstecki2015}.  Pushing the flow rate to even higher values yields the fragmentation of the oil-water interfaces, and to the formation of both direct and inverse emulsions. 

These four dynamical patterns have a clear signature on four global observables. In Fig.~\ref{fourregimes}b we show the variations of the water saturation $S_{\rm w}$ defined as the volume fraction of water, or equivalently as the fraction of oil extracted from the porous medium, when the water front reaches the end of the device.  $S_{\rm w}$ undergoes non-monotonic variations with $Ca$ and reaches a minimum exactly at the transition between regimes B and C. Surprisingly, at small flow rates pushing faster hinders oil extraction. Ignoring the heterogeneities of the water-film thickness,  the area fraction of the porous medium in contact with water, $A_{\rm w}$, also varies non-monotonically with $Ca$. $A_{\rm w}$ reaches a minimum when a river network emerges (regime C). Above this transition, the increase of $A_{\rm w}$ echoes the widening of  the rivers. 

\begin{figure*}
\begin{center}
  \includegraphics[width=\textwidth]{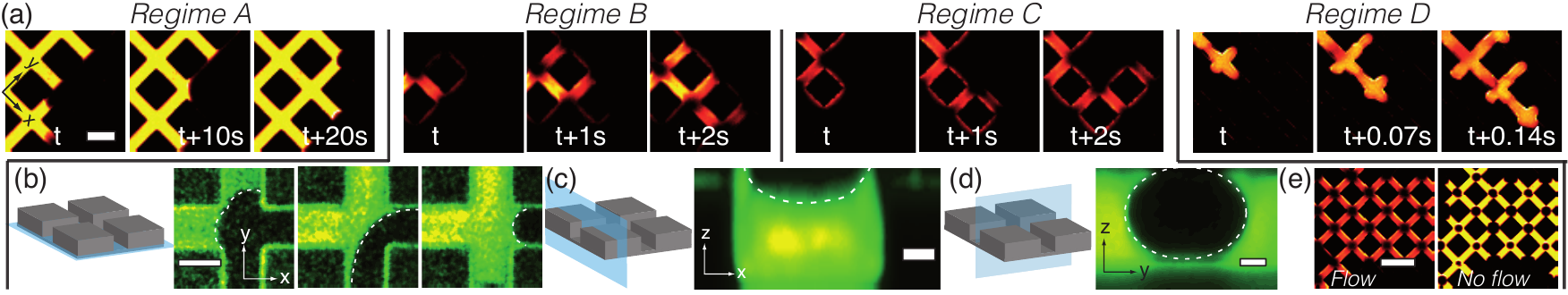}
  \caption{\label{dynamics}
  (a): Consecutive snapshots of the interface dynamics at the pore scale in the four regimes. In regime B, the thin films are unstable, oil droplets  form and are trapped at the vertices. In regime C the thin films are stable.  Note the reversal of the apparent contact angle for regime D. Same Capillary numbers as in Fig.~\ref{fourregimes}a. { Scale bar: $200\,\rm \mu m$}. 
(b) Subsequent confocal images in the $xy$ plane close to the sticker surface as sketched in the left panel. The contact line progresses everywhere past the sticker surface. {Scale bar: $100\,\rm \mu m$}. (c) Confocal image of the cross section of a channel in regime B. Note that the water film is localized on the bottom surface, leaving an oil hemicylinder on the upper wall. {Scale bar: $20\,\rm \mu m$}. (d) Same experiment imaged in the $yz$ plane at the intersection between two channels: an oil droplet is trapped. Scale bar: $20\,\rm \mu m$. (e) Destabilization of the thin films (Regime C). Left picture: water flows and the films are stable ($Ca=2.2\times 10^{-4}$). Right picture: the flow is stopped and the films are destabilized leaving oil droplets at the channel intersections {Scale bar: $400\,\rm \mu m$}.
  }
\end{center}
\end{figure*}

The transitions toward the  two other regimes is clearly visible when inspecting  the distribution of the water-film thickness, $P(h)$, and the non-monotonic variations of $\langle h\rangle$  in Figs.~\ref{fourregimes}d and \ref{fourregimes}e. At low $Ca$ (regime A), the distribution is bimodal: the global maximum is located at $h=H$, the mean height of the channels. A smaller peak also exists at $h=0$ and trivially corresponds to the boundaries of the imbibition pattern. 
The transition from regime A to regime B coincides with the inversion of the magnitude of the peaks at $h=0$ and $h=H$  in Fig.~\ref{fourregimes}d.  The ratio  between the typical and the average film thickness drops down to zero at this transition,  Fig. \ref{fourregimes}e.  The second transition from regimes B to C is also visible on $P(h)$. When the river network forms, the thickness distribution is peaked at $h=0$ and plateaus at intermediate thicknesses. Finally,  the  increase of the mean thickness $\langle h \rangle$ in Fig.~\ref{fourregimes}e signals the onset of regime D.

Three comments are in order. Firstly, the abrupt changes in these macroscopic observables confirm that the four imbibition regimes correspond to four distinct physical processes. Secondly, as demonstrated in a Supplementary Document~\cite{supp}, the same transitions from regimes A to B, and B to C are also observed keeping $Ca$ constant and increasing the viscosity ratio $M$.  Consistently, these transitions  have the same signature on the film-thickness distribution.  Finally, this classification in four regimes  extends that proposed in~\cite{Juanes1} where disorder and the radial variations of the local flow rate are likely to blur the transitions between  regimes A, B, and C.

We now single out the  origin of these four regimes by investigating the imbibition dynamics at the pore scale illustrated in  Fig.~\ref{dynamics}a. 

\noindent {\em -- Regime A:}  The  corresponding snapshots  provide  a prototypical illustration of capillary imbibition. The menisci proceed uniformly through the channels, merge at the vertices and grow until touching the next post. Then, they split into two interfaces  invading the adjacent channels. 
\begin{figure*}
\begin{center}
  \includegraphics[width=\textwidth]{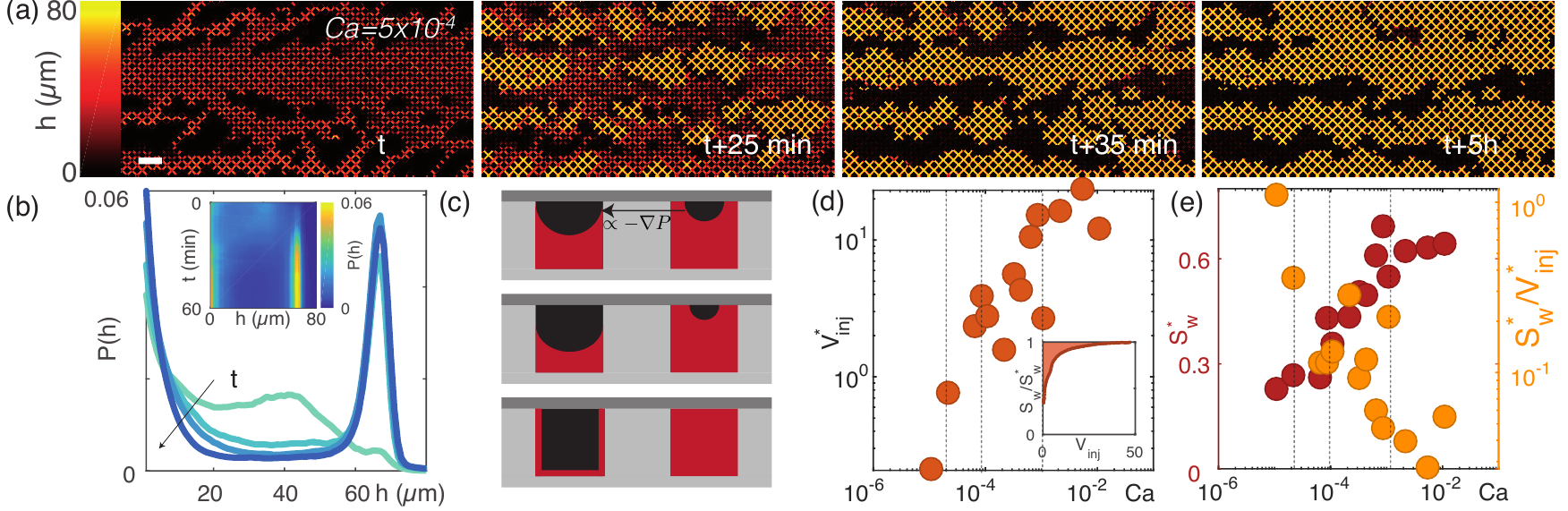}
  \caption{\label{coarsening}
  Coarsening dynamics. (a) Snapshot of the imbibition films in regime C, where the color indicates the local height of the water films. The first snapshot is taken when water percolates through the lattice, $t\sim 150\,\rm s$ after injection. The thin films coarsen. $Ca= 5 \times 10^{-4}$, $M=560$, scale bar: $1\,\rm mm$. (b) Distribution of the water-film thickness at  times corresponding to the snapshots shown in (a). Inset: full temporal evolution of the film-thickness distribution. (c) Sketch of the coarsening mechanism. Oil flows from the regions of highest Laplace pressure. Equivalently, mass conservation implies that the thinner water films empty in the thicker. { (d) Variation of $V^{\star}_{\rm inj}$ with the capillary number. Inset: Measure of $V^{\star}_{\rm inj}$. The typical injection volume to reach a steady state corresponds to the area above the curve $S_{\rm w}(V_{\rm inj})/S_{\rm w}^\star$. (e) Opposite variations of $S^{\star}_{\rm w}$ and $S^{\star}_{\rm w}/V^{\star}_{\rm inj}$ with the capillary number.}
  }
\end{center}
\end{figure*}

\noindent {\em -- Regime B and C:} The   corresponding snapshot sequences reveal a marked change in the microscopic  dynamics. In both regimes, as the contact line reaches a corner formed by a square post and a confining wall, the perimeter of the post is coated leaving  a significant fraction of oil immobile in  the channels.  The coating layer then growths, contacts adjacent posts, and therefore discretely progresses through the  lattice. This behavior is consistent with the observations of~\cite{Juanes1}. Confocal imaging is however necessary to complete this   scenario. The confocal pictures  in  Fig.~\ref{dynamics}b are taken $5\,\rm \mu m$ above the bottom surface of the channel (sticker surface).  The progression of the water-oil interface in this plane is very similar to the capillary-imbibition sequences illustrating regime A in  Fig.~\ref{dynamics}a.  Even though the  contact line accelerates when  reaching a corner, it also freely advances away from the posts along the direction of the microchannels. The advancing of the water film does not require any additional Laplace-pressure pumping from the negatively curved gutters at the base of the wetted posts~\cite{Concus69}. {A possible explanation could be the existence of a rough structure further driving the imbibition process at the sub-optical scales.} This hypothesis is ruled out by Atomic Force Microscopy imaging, see Supplementary Informations. Both the quartz and sticker surfaces are smooth down to the $10\,\rm nm$ scale~\cite{supp}.  Therefore the only possible  reason for the formation of a thin film in regimes B and C is a dynamical wetting transition~\cite{Andreotti2013,Levache2014}. 
The confocal image in the $xz$ plane indeed shows that  water films  are entrained only on the bottom surface and fully cover it as they grow, Fig.~\ref{dynamics}c. Increasing the capillary number, the contact line on the sticker surface entrains the aqueous liquid at higher speed than the average liquid front. The localization of the entrainment  on the sticker surface   is very likely to stem from the  lower advancing contact angles value. Liquid entrainment  prevents the steady motion of a stable meniscus in agreement with observations in simpler Hele-Shaw geometries~\cite{Levache2014}. 

The  dynamics of the water front are identical in regimes B and C.  However an interfacial instability distinguishes these two regimes, characterized by  different  film thickness distributions in Fig.~\ref{fourregimes}d. In both cases, the incomplete displacement of the non-wetting fluid results in the formation of hemicylindrical interfaces, Fig.~\ref{dynamics}c. In regime B, these interfaces are unstable to the Rayleigh-Plateau instability which breaks the oil cylinders into droplets receding to the vertices of the lattice, Fig.~\ref{dynamics}a and ~\ref{dynamics}d. 
%
{Conversely, increasing the flow rate to regime C, the Rayleigh-Plateau instability is suppressed. In principle the stabilization of the  oil hemicylinders could be either due to local pinning of the contact line~\cite{Lipowsky99,Lauga2009}, or by the convection of the interface fluctuations as in quickly stretched liquid filaments~\cite{Tomotika}, and in coflowing liquid streams~\cite{Villermaux}.
In order to single out the origin of stabilization, we performed additional experiments deep in regime C ($Ca=2.2\times 10^{-4}$) followed by an abrupt stop of the water flow illustrated in fig.~\ref{dynamics}e. As the flow  stops, the interfaces become unstable. The snap-off events and  the resulting  droplet pattern typical of regime B are recovered. This observation unambiguously confirms that the oil interface is dynamically stabilized by the flows in the water films.}

\noindent {\em -- Regime D}: At the highest capillary numbers, the local dynamics undergoes a strong qualitative change. Given the speed of the water front, confocal imaging is not possible anymore. However, it is already clear from the high-resolution snapshots in Fig.~\ref{dynamics}a that the apparent contact angle between the aqueous liquid and the solid walls is reversed: the porous medium becomes effectively hydrophobic. We explain this reversal  by another wetting transition first reported for colloidal liquids with  ultra-low surface tensions~\cite{Aarts2013}. In this high-$Ca$ regime the destabilization of the interface occurs at the center of the channels where a finger of the aqueous liquid grows and leaves an immobile oil layer on the solid surfaces. In turn, the lubricated motion of the water fingers drastically changes  the exploration of the post lattice. The imbibition patterns become analogous to dendritic growth as observed in conventional anisotropic drainage experiments~\cite{Chen85,Sander85,Garstecki2015}. 

In line with earlier studies, until now we have  limited our discussion to pattern formation prior to the percolation of the driving fluid through the lattice. This limitation is unimportant when  liquid mobilization is complete at the pore scale (i.e. in drainage experiments and in regime A). The driving fluid then merely flows through the percolating pattern which  does not age~\cite{Lenormand90}. In  contrast,  we now demonstrate that the thin-film patterns undergo significant structural changes over long time scales.
Fig.~\ref{coarsening}a show the evolution of the film-thickness distribution over hours of continuous injection in regime C ($Ca=5 \times 10^{-4}$). From now on we measure time in term of the volume of injected water  normalized by the overall volume of the porous medium  $V_{\rm inj}=Qt/V_{\rm Total}$. In the example of Fig.~\ref{coarsening}a, $V_{\rm inj}=0.27$ at percolation, and the pattern forms a river network. Further increasing the injected volume, the oil-water interfaces are not destabilized by the Rayleigh-Plateau instability. Instead the wetting pattern slowly evolves into compact islands of pores fully filled with water connected by a sea of thin films, Fig.~\ref{coarsening}a. The islands grow  at the expense of the thin films and eventually coalesce into a percolating cluster of pores where oil has been fully displaced. 
More quantitatively, see Figs.~\ref{coarsening}b and \ref{coarsening}b inset, the water-thickness distribution  evolves from a decreasing function with a thickness plateau comprised between 10 and 50 microns, to a bimodal distribution peaked both at $h=0$ (pattern edges and residual thin films) and $h=H=70\,\rm \mu m$ (filled channels). This slow dynamics has a clear microscopic origin. As sketched in Fig.~\ref{coarsening}c, the channels where the water films are thin correspond to smaller curvature of the oil-water interface, and hence to smaller Laplace pressure.  Therefore Laplace pressure gradients cause the oil to flow out of  regions where the water-films are thicker until total water saturation. This coarsening process relies on the capillary drive of the more viscous liquids which {\em de facto} results in a slower  dynamics than the forced propagation of the water films. The collective relaxation of the initial  distribution of $h$ further increases the volume of injected water needed  to reach a stationary pattern. The typical water volume, $V_{\rm inj}^\star$ required to reach the asymptotic steady state is measured from the variations of the the water saturation $S_{\rm w}$, Fig.~\ref{coarsening}d inset.  We find in Fig.~\ref{coarsening}d  that $V_{\rm inj}^\star$ monotonically increases with $Ca$: the higher the flow rate the higher the water volume required to achieve  maximal saturation and form stationary patterns. However, unlike its value at  percolation, the asymptotic value of $S_{\rm w}^\star$ monotonically increases with $Ca$: pushing harder mobilizes more oil, see Fig.~\ref{coarsening}e.

Together  the variations of $V_{\rm inj}^\star$, and $S_{\rm w}^\star$ are useful guidelines to design optimal oil-recovery strategies. Decontamination processes requiring the mobilization of the maximal amount of oil, at any cost, would be optimized at the highest possible injection rate. Conversely,  commercial oil extraction would require a trade-off, e.g. maximizing the $S_{\rm w}^\star/V_{\rm inj}^\star$ ratio, which would correspond to minimizing the water flow rate, Fig.~\ref{coarsening}e (light circles). From a more fundamental perspective, we hope our findings will stimulate  experimental and theoretical investigation to quantitatively elucidate the interfacial instabilities underlying the fourfold dynamical scenarios established in this letter.


\acknowledgements
We acknowledge support from Institut Universitaire de France (D. B.). We thank M. Faivre, C.Moscalenko and M. Levant for help with the lithography,  AFM and SEM experiments. We also thank R. Juanes, G. Mckinley and  A. Pahlavan for insightful comments and valuable  suggestions.

\begin{thebibliography}{100}
\makeatletter
\providecommand \@ifxundefined [1]{%
 \@ifx{#1\undefined}
}%
\providecommand \@ifnum [1]{%
 \ifnum #1\expandafter \@firstoftwo
 \else \expandafter \@secondoftwo
 \fi
}%
\providecommand \@ifx [1]{%
 \ifx #1\expandafter \@firstoftwo
 \else \expandafter \@secondoftwo
 \fi
}%
\providecommand \natexlab [1]{#1}%
\providecommand \enquote  [1]{``#1''}%
\providecommand \bibnamefont  [1]{#1}%
\providecommand \bibfnamefont [1]{#1}%
\providecommand \citenamefont [1]{#1}%
\providecommand \href@noop [0]{\@secondoftwo}%
\providecommand \href [0]{\begingroup \@sanitize@url \@href}%
\providecommand \@href[1]{\@@startlink{#1}\@@href}%
\providecommand \@@href[1]{\endgroup#1\@@endlink}%
\providecommand \@sanitize@url [0]{\catcode `\\12\catcode `\$12\catcode
  `\&12\catcode `\#12\catcode `\^12\catcode `\_12\catcode `\%12\relax}%
\providecommand \@@startlink[1]{}%
\providecommand \@@endlink[0]{}%
\providecommand \url  [0]{\begingroup\@sanitize@url \@url }%
\providecommand \@url [1]{\endgroup\@href {#1}{\urlprefix }}%
\providecommand \urlprefix  [0]{URL }%
\providecommand \Eprint [0]{\href }%
\providecommand \doibase [0]{http://dx.doi.org/}%
\providecommand \selectlanguage [0]{\@gobble}%
\providecommand \bibinfo  [0]{\@secondoftwo}%
\providecommand \bibfield  [0]{\@secondoftwo}%
\providecommand \translation [1]{[#1]}%
\providecommand \BibitemOpen [0]{}%
\providecommand \bibitemStop [0]{}%
\providecommand \bibitemNoStop [0]{.\EOS\space}%
\providecommand \EOS [0]{\spacefactor3000\relax}%
\providecommand \BibitemShut  [1]{\csname bibitem#1\endcsname}%
\let\auto@bib@innerbib\@empty
\bibitem [{\citenamefont {Buckley}\ and\ \citenamefont
  {Leverett}(1942)}]{Buckley42}%
  \BibitemOpen
  \bibfield  {author} {\bibinfo {author} {\bibfnamefont {S.~E.}\ \bibnamefont
  {Buckley}}\ and\ \bibinfo {author} {\bibfnamefont {M.~C.}\ \bibnamefont
  {Leverett}},\ }\bibfield  {title} {\enquote {\bibinfo {title} {Mechanism of
  fluid displacement in sands},}\ }\href
  {https://www.onepetro.org/journal-paper/SPE-942107-G} {\bibfield  {journal}
  {\bibinfo  {journal} {Transactions of the American Institute of Mining and
  Metallurgical Engineers}\ }\textbf {\bibinfo {volume} {146}},\ \bibinfo
  {pages} {107--116} (\bibinfo {year} {1942})}\BibitemShut {NoStop}%
\bibitem [{\citenamefont {Lenormand}\ \emph {et~al.}(1988)\citenamefont
  {Lenormand}, \citenamefont {Touboul},\ and\ \citenamefont
  {Zarcone}}]{Lenormand1897}%
  \BibitemOpen
  \bibfield  {author} {\bibinfo {author} {\bibfnamefont {Roland}\ \bibnamefont
  {Lenormand}}, \bibinfo {author} {\bibfnamefont {Eric}\ \bibnamefont
  {Touboul}}, \ and\ \bibinfo {author} {\bibfnamefont {Cesar}\ \bibnamefont
  {Zarcone}},\ }\bibfield  {title} {\enquote {\bibinfo {title} {Numerical
  models and experiments on immiscible displacements in porous media},}\ }\href
  {https://www.cambridge.org/core/journals/journal-of-fluid-mechanics/article/div-classtitlenumerical-models-and-experiments-on-immiscible-displacements-in-porous-mediadiv/E5D96D976F98B7670CCF3C64617A262E}
  {\bibfield  {journal} {\bibinfo  {journal} {Journal of Fluid Mechanics}\
  }\textbf {\bibinfo {volume} {189}},\ \bibinfo {pages} {165?187} (\bibinfo
  {year} {1988})}\BibitemShut {NoStop}%
\bibitem [{\citenamefont {Lenormand}(1990)}]{Lenormand90}%
  \BibitemOpen
  \bibfield  {author} {\bibinfo {author} {\bibfnamefont {R.}~\bibnamefont
  {Lenormand}},\ }\bibfield  {title} {\enquote {\bibinfo {title} {Liquids in
  porous media},}\ }\href {http://stacks.iop.org/0953-8984/2/i=S/a=008}
  {\bibfield  {journal} {\bibinfo  {journal} {Journal of Physics: Condensed
  Matter}\ }\textbf {\bibinfo {volume} {2}},\ \bibinfo {pages} {SA79} (\bibinfo
  {year} {1990})}\BibitemShut {NoStop}%
\bibitem [{\citenamefont {Sandnes}\ \emph {et~al.}(2011)\citenamefont
  {Sandnes}, \citenamefont {Flekk{\o}y}, \citenamefont {Knudsen}, \citenamefont
  {M{\aa}l{\o}y},\ and\ \citenamefont {See}}]{Sandnes2011}%
  \BibitemOpen
  \bibfield  {author} {\bibinfo {author} {\bibfnamefont {B.}~\bibnamefont
  {Sandnes}}, \bibinfo {author} {\bibfnamefont {E.~G.}\ \bibnamefont
  {Flekk{\o}y}}, \bibinfo {author} {\bibfnamefont {H.~A.}\ \bibnamefont
  {Knudsen}}, \bibinfo {author} {\bibfnamefont {K.~J.}\ \bibnamefont
  {M{\aa}l{\o}y}}, \ and\ \bibinfo {author} {\bibfnamefont {H.}~\bibnamefont
  {See}},\ }\bibfield  {title} {\enquote {\bibinfo {title} {Patterns and flow
  in frictional fluid dynamics},}\ }\href
  {http://www.nature.com/articles/ncomms1289} {\bibfield  {journal} {\bibinfo
  {journal} {Nature Communications}\ }\textbf {\bibinfo {volume} {2}},\
  \bibinfo {pages} {288 EP --} (\bibinfo {year} {2011})}\BibitemShut {NoStop}%
\bibitem [{\citenamefont {Sahimi}(1993)}]{Sahimi93}%
  \BibitemOpen
  \bibfield  {author} {\bibinfo {author} {\bibfnamefont {Muhammad}\
  \bibnamefont {Sahimi}},\ }\bibfield  {title} {\enquote {\bibinfo {title}
  {Flow phenomena in rocks: from continuum models to fractals, percolation,
  cellular automata, and simulated annealing},}\ }\href
  {http://journals.aps.org/rmp/abstract/10.1103/RevModPhys.65.1393} {\bibfield
  {journal} {\bibinfo  {journal} {Rev. Mod. Phys.}\ }\textbf {\bibinfo {volume}
  {65}},\ \bibinfo {pages} {1393--1534} (\bibinfo {year} {1993})}\BibitemShut
  {NoStop}%
\bibitem [{\citenamefont {Alava}\ \emph {et~al.}(2004)\citenamefont {Alava},
  \citenamefont {Dube},\ and\ \citenamefont {Rost}}]{Alava2004}%
  \BibitemOpen
  \bibfield  {author} {\bibinfo {author} {\bibfnamefont {M.}~\bibnamefont
  {Alava}}, \bibinfo {author} {\bibfnamefont {M.}~\bibnamefont {Dube}}, \ and\
  \bibinfo {author} {\bibfnamefont {M.}~\bibnamefont {Rost}},\ }\bibfield
  {title} {\enquote {\bibinfo {title} {Imbibition in disordered media},}\
  }\href {http://www.tandfonline.com/doi/abs/10.1080/00018730410001687363}
  {\bibfield  {journal} {\bibinfo  {journal} {Advances in Physics}\ }\textbf
  {\bibinfo {volume} {53}},\ \bibinfo {pages} {83--175} (\bibinfo {year}
  {2004})}\BibitemShut {NoStop}%
\bibitem [{\citenamefont {Stokes}\ \emph {et~al.}(1986)\citenamefont {Stokes},
  \citenamefont {Weitz}, \citenamefont {Gollub}, \citenamefont {Dougherty},
  \citenamefont {Robbins}, \citenamefont {Chaikin},\ and\ \citenamefont
  {Lindsay}}]{Chaikin86}%
  \BibitemOpen
  \bibfield  {author} {\bibinfo {author} {\bibfnamefont {J.~P.}\ \bibnamefont
  {Stokes}}, \bibinfo {author} {\bibfnamefont {D.~A.}\ \bibnamefont {Weitz}},
  \bibinfo {author} {\bibfnamefont {J.~P.}\ \bibnamefont {Gollub}}, \bibinfo
  {author} {\bibfnamefont {A.}~\bibnamefont {Dougherty}}, \bibinfo {author}
  {\bibfnamefont {M.~O.}\ \bibnamefont {Robbins}}, \bibinfo {author}
  {\bibfnamefont {P.~M.}\ \bibnamefont {Chaikin}}, \ and\ \bibinfo {author}
  {\bibfnamefont {H.~M.}\ \bibnamefont {Lindsay}},\ }\bibfield  {title}
  {\enquote {\bibinfo {title} {Interfacial stability of immiscible displacement
  in a porous medium},}\ }\href
  {http://journals.aps.org/prl/abstract/10.1103/PhysRevLett.57.1718} {\bibfield
   {journal} {\bibinfo  {journal} {Phys. Rev. Lett.}\ }\textbf {\bibinfo
  {volume} {57}},\ \bibinfo {pages} {1718--1721} (\bibinfo {year}
  {1986})}\BibitemShut {NoStop}%
\bibitem [{\citenamefont {Weitz}\ \emph {et~al.}(1987)\citenamefont {Weitz},
  \citenamefont {Stokes}, \citenamefont {Ball},\ and\ \citenamefont
  {Kushnick}}]{Stokes87}%
  \BibitemOpen
  \bibfield  {author} {\bibinfo {author} {\bibfnamefont {D.~A.}\ \bibnamefont
  {Weitz}}, \bibinfo {author} {\bibfnamefont {J.~P.}\ \bibnamefont {Stokes}},
  \bibinfo {author} {\bibfnamefont {R.~C.}\ \bibnamefont {Ball}}, \ and\
  \bibinfo {author} {\bibfnamefont {A.~P.}\ \bibnamefont {Kushnick}},\
  }\bibfield  {title} {\enquote {\bibinfo {title} {Dynamic capillary pressure
  in porous media: Origin of the viscous-fingering length scale},}\ }\href
  {\doibase 10.1103/PhysRevLett.59.2967} {\bibfield  {journal} {\bibinfo
  {journal} {Phys. Rev. Lett.}\ }\textbf {\bibinfo {volume} {59}},\ \bibinfo
  {pages} {2967--2970} (\bibinfo {year} {1987})}\BibitemShut {NoStop}%
\bibitem [{\citenamefont {Levach\'e}\ and\ \citenamefont
  {Bartolo}(2014)}]{Levache2014}%
  \BibitemOpen
  \bibfield  {author} {\bibinfo {author} {\bibfnamefont {Bertrand}\
  \bibnamefont {Levach\'e}}\ and\ \bibinfo {author} {\bibfnamefont {Denis}\
  \bibnamefont {Bartolo}},\ }\bibfield  {title} {\enquote {\bibinfo {title}
  {Revisiting the saffman-taylor experiment: Imbibition patterns and
  liquid-entrainment transitions},}\ }\href
  {http://journals.aps.org/prl/abstract/10.1103/PhysRevLett.113.044501}
  {\bibfield  {journal} {\bibinfo  {journal} {Phys. Rev. Lett.}\ }\textbf
  {\bibinfo {volume} {113}},\ \bibinfo {pages} {044501} (\bibinfo {year}
  {2014})}\BibitemShut {NoStop}%
\bibitem [{\citenamefont {Wexler}\ \emph {et~al.}(2015)\citenamefont {Wexler},
  \citenamefont {Jacobi},\ and\ \citenamefont {Stone}}]{Wexler1}%
  \BibitemOpen
  \bibfield  {author} {\bibinfo {author} {\bibfnamefont {Jason~S.}\
  \bibnamefont {Wexler}}, \bibinfo {author} {\bibfnamefont {Ian}\ \bibnamefont
  {Jacobi}}, \ and\ \bibinfo {author} {\bibfnamefont {Howard~A.}\ \bibnamefont
  {Stone}},\ }\bibfield  {title} {\enquote {\bibinfo {title} {Shear-driven
  failure of liquid-infused surfaces},}\ }\href {\doibase
  10.1103/PhysRevLett.114.168301} {\bibfield  {journal} {\bibinfo  {journal}
  {Phys. Rev. Lett.}\ }\textbf {\bibinfo {volume} {114}},\ \bibinfo {pages}
  {168301} (\bibinfo {year} {2015})}\BibitemShut {NoStop}%
\bibitem [{\citenamefont {Zhao}\ \emph {et~al.}(2016)\citenamefont {Zhao},
  \citenamefont {MacMinn},\ and\ \citenamefont {Juanes}}]{Juanes1}%
  \BibitemOpen
  \bibfield  {author} {\bibinfo {author} {\bibfnamefont {Benzhong}\
  \bibnamefont {Zhao}}, \bibinfo {author} {\bibfnamefont {Christopher~W.}\
  \bibnamefont {MacMinn}}, \ and\ \bibinfo {author} {\bibfnamefont {Ruben}\
  \bibnamefont {Juanes}},\ }\bibfield  {title} {\enquote {\bibinfo {title}
  {Wettability control on multiphase flow in patterned microfluidics},}\ }\href
  {http://www.pnas.org/content/113/37/10251.full} {\bibfield  {journal}
  {\bibinfo  {journal} {Proceedings of the National Academy of Sciences}\
  }\textbf {\bibinfo {volume} {113}},\ \bibinfo {pages} {10251--10256}
  (\bibinfo {year} {2016})}\BibitemShut {NoStop}%
\bibitem [{\citenamefont {Trojer}\ \emph {et~al.}(2015)\citenamefont {Trojer},
  \citenamefont {Szulczewski},\ and\ \citenamefont {Juanes}}]{Juanes2}%
  \BibitemOpen
  \bibfield  {author} {\bibinfo {author} {\bibfnamefont {Mathias}\ \bibnamefont
  {Trojer}}, \bibinfo {author} {\bibfnamefont {Michael~L.}\ \bibnamefont
  {Szulczewski}}, \ and\ \bibinfo {author} {\bibfnamefont {Ruben}\ \bibnamefont
  {Juanes}},\ }\bibfield  {title} {\enquote {\bibinfo {title} {Stabilizing
  fluid-fluid displacements in porous media through wettability alteration},}\
  }\href
  {http://journals.aps.org/prapplied/abstract/10.1103/PhysRevApplied.3.054008}
  {\bibfield  {journal} {\bibinfo  {journal} {Phys. Rev. Applied}\ }\textbf
  {\bibinfo {volume} {3}},\ \bibinfo {pages} {054008} (\bibinfo {year}
  {2015})}\BibitemShut {NoStop}%
\bibitem [{\citenamefont {Holtzman}\ and\ \citenamefont
  {Segre}(2015)}]{Segre2015}%
  \BibitemOpen
  \bibfield  {author} {\bibinfo {author} {\bibfnamefont {Ran}\ \bibnamefont
  {Holtzman}}\ and\ \bibinfo {author} {\bibfnamefont {Enrico}\ \bibnamefont
  {Segre}},\ }\bibfield  {title} {\enquote {\bibinfo {title} {Wettability
  stabilizes fluid invasion into porous media via nonlocal, cooperative pore
  filling},}\ }\href {\doibase 10.1103/PhysRevLett.115.164501} {\bibfield
  {journal} {\bibinfo  {journal} {Phys. Rev. Lett.}\ }\textbf {\bibinfo
  {volume} {115}},\ \bibinfo {pages} {164501} (\bibinfo {year}
  {2015})}\BibitemShut {NoStop}%
  %
 \bibitem{supp}
See Supplementary Informations where we report: (i) a detailed description of the materials and methods, (ii) an AFM characterization of the surface roughness and (iii) experiments performed at constant flow rate and increasing viscosity ratio.
%
\bibitem [{\citenamefont {Bartolo}\ \emph {et~al.}(2008)\citenamefont
  {Bartolo}, \citenamefont {Degre}, \citenamefont {Nghe},\ and\ \citenamefont
  {Studer}}]{Bartolo2008}%
  \BibitemOpen
  \bibfield  {author} {\bibinfo {author} {\bibfnamefont {Denis}\ \bibnamefont
  {Bartolo}}, \bibinfo {author} {\bibfnamefont {Guillaume}\ \bibnamefont
  {Degre}}, \bibinfo {author} {\bibfnamefont {Philippe}\ \bibnamefont {Nghe}},
  \ and\ \bibinfo {author} {\bibfnamefont {Vincent}\ \bibnamefont {Studer}},\
  }\bibfield  {title} {\enquote {\bibinfo {title} {Microfluidic stickers},}\
  }\href
  {http://pubs.rsc.org/en/Content/ArticleLanding/2008/LC/B712368J#!divAbstract}
  {\bibfield  {journal} {\bibinfo  {journal} {Lab Chip}\ }\textbf {\bibinfo
  {volume} {8}},\ \bibinfo {pages} {274--279} (\bibinfo {year}
  {2008})}\BibitemShut {NoStop}%
  %
\bibitem [{\citenamefont {Levache}\ \emph {et~al.}(2012)\citenamefont
  {Levache}, \citenamefont {Azioune}, \citenamefont {Bourrel}, \citenamefont
  {Studer},\ and\ \citenamefont {Bartolo}}]{Levache2012}%
  \BibitemOpen
  \bibfield  {author} {\bibinfo {author} {\bibfnamefont {Bertrand}\
  \bibnamefont {Levache}}, \bibinfo {author} {\bibfnamefont {Ammar}\
  \bibnamefont {Azioune}}, \bibinfo {author} {\bibfnamefont {Maurice}\
  \bibnamefont {Bourrel}}, \bibinfo {author} {\bibfnamefont {Vincent}\
  \bibnamefont {Studer}}, \ and\ \bibinfo {author} {\bibfnamefont {Denis}\
  \bibnamefont {Bartolo}},\ }\bibfield  {title} {\enquote {\bibinfo {title}
  {Engineering the surface properties of microfluidic stickers},}\ }\href
  {http://pubs.rsc.org/en/content/articlelanding/2012/lc/c2lc40284j#!divAbstract}
  {\bibfield  {journal} {\bibinfo  {journal} {Lab Chip}\ }\textbf {\bibinfo
  {volume} {12}},\ \bibinfo {pages} {3028--3031} (\bibinfo {year}
  {2012})}\BibitemShut {NoStop}%
\bibitem [{\citenamefont {Courbin}\ \emph {et~al.}(2007)\citenamefont
  {Courbin}, \citenamefont {Denieul}, \citenamefont {Dressaire}, \citenamefont
  {Roper}, \citenamefont {Ajdari},\ and\ \citenamefont {Stone}}]{Courbin2007}%
  \BibitemOpen
  \bibfield  {author} {\bibinfo {author} {\bibfnamefont {Laurent}\ \bibnamefont
  {Courbin}}, \bibinfo {author} {\bibfnamefont {Etienne}\ \bibnamefont
  {Denieul}}, \bibinfo {author} {\bibfnamefont {Emilie}\ \bibnamefont
  {Dressaire}}, \bibinfo {author} {\bibfnamefont {Marcus}\ \bibnamefont
  {Roper}}, \bibinfo {author} {\bibfnamefont {Armand}\ \bibnamefont {Ajdari}},
  \ and\ \bibinfo {author} {\bibfnamefont {Howard~A.}\ \bibnamefont {Stone}},\
  }\bibfield  {title} {\enquote {\bibinfo {title} {Imbibition by polygonal
  spreading on microdecorated surfaces},}\ }\href
  {http://www.nature.com/nmat/journal/v6/n9/full/nmat1978.html} {\bibfield
  {journal} {\bibinfo  {journal} {Nature Materials}\ ,\ \bibinfo {pages}
  {661--664}} (\bibinfo {year} {2007})}\BibitemShut {NoStop}%
\bibitem [{\citenamefont {Cubaud}\ and\ \citenamefont
  {Fermigier}(2001)}]{Cubaud2001}%
  \BibitemOpen
  \bibfield  {author} {\bibinfo {author} {\bibfnamefont {T.}~\bibnamefont
  {Cubaud}}\ and\ \bibinfo {author} {\bibfnamefont {M.}~\bibnamefont
  {Fermigier}},\ }\bibfield  {title} {\enquote {\bibinfo {title} {Faceted drops
  on heterogeneous surfaces},}\ }\href
  {http://iopscience.iop.org/article/10.1209/epl/i2001-00405-1/meta} {\bibfield
   {journal} {\bibinfo  {journal} {EPL (Europhysics Letters)}\ }\textbf
  {\bibinfo {volume} {55}},\ \bibinfo {pages} {239} (\bibinfo {year}
  {2001})}\BibitemShut {NoStop}%
\bibitem [{\citenamefont {Chen}\ and\ \citenamefont
  {Wilkinson}(1985)}]{Chen85}%
  \BibitemOpen
  \bibfield  {author} {\bibinfo {author} {\bibfnamefont {Jing-Den}\
  \bibnamefont {Chen}}\ and\ \bibinfo {author} {\bibfnamefont {David}\
  \bibnamefont {Wilkinson}},\ }\bibfield  {title} {\enquote {\bibinfo {title}
  {Pore-scale viscous fingering in porous media},}\ }\href
  {http://journals.aps.org/prl/abstract/10.1103/PhysRevLett.55.1892} {\bibfield
   {journal} {\bibinfo  {journal} {Phys. Rev. Lett.}\ }\textbf {\bibinfo
  {volume} {55}},\ \bibinfo {pages} {1892--1895} (\bibinfo {year}
  {1985})}\BibitemShut {NoStop}%
\bibitem [{\citenamefont {Ben-Jacob}\ \emph {et~al.}(1985)\citenamefont
  {Ben-Jacob}, \citenamefont {Godbey}, \citenamefont {Goldenfeld},
  \citenamefont {Koplik}, \citenamefont {Levine}, \citenamefont {Mueller},\
  and\ \citenamefont {Sander}}]{Sander85}%
  \BibitemOpen
  \bibfield  {author} {\bibinfo {author} {\bibfnamefont {E.}~\bibnamefont
  {Ben-Jacob}}, \bibinfo {author} {\bibfnamefont {R.}~\bibnamefont {Godbey}},
  \bibinfo {author} {\bibfnamefont {Nigel~D.}\ \bibnamefont {Goldenfeld}},
  \bibinfo {author} {\bibfnamefont {J.}~\bibnamefont {Koplik}}, \bibinfo
  {author} {\bibfnamefont {H.}~\bibnamefont {Levine}}, \bibinfo {author}
  {\bibfnamefont {T.}~\bibnamefont {Mueller}}, \ and\ \bibinfo {author}
  {\bibfnamefont {L.~M.}\ \bibnamefont {Sander}},\ }\bibfield  {title}
  {\enquote {\bibinfo {title} {Experimental demonstration of the role of
  anisotropy in interfacial pattern formation},}\ }\href
  {http://journals.aps.org/prl/abstract/10.1103/PhysRevLett.55.1315} {\bibfield
   {journal} {\bibinfo  {journal} {Phys. Rev. Lett.}\ }\textbf {\bibinfo
  {volume} {55}},\ \bibinfo {pages} {1315--1318} (\bibinfo {year}
  {1985})}\BibitemShut {NoStop}%
\bibitem [{\citenamefont {Budek}\ \emph {et~al.}({2015})\citenamefont {Budek},
  \citenamefont {Garstecki}, \citenamefont {Samborski},\ and\ \citenamefont
  {Szymczak}}]{Garstecki2015}%
  \BibitemOpen
  \bibfield  {author} {\bibinfo {author} {\bibfnamefont {Agnieszka}\
  \bibnamefont {Budek}}, \bibinfo {author} {\bibfnamefont {Piotr}\ \bibnamefont
  {Garstecki}}, \bibinfo {author} {\bibfnamefont {Adam}\ \bibnamefont
  {Samborski}}, \ and\ \bibinfo {author} {\bibfnamefont {Piotr}\ \bibnamefont
  {Szymczak}},\ }\bibfield  {title} {\enquote {\bibinfo {title} {{Thin-finger
  growth and droplet pinch-off in miscible and immiscible displacements in a
  periodic network of microfluidic channels}},}\ }\href
  {http://aip.scitation.org/doi/abs/10.1063/1.4935225?journalCode=phf}
  {\bibfield  {journal} {\bibinfo  {journal} {{Physics of Fluids}}\ }\textbf
  {\bibinfo {volume} {{27}}} (\bibinfo {year} {{2015}})}\BibitemShut {NoStop}%
\bibitem [{\citenamefont {Concus}\ and\ \citenamefont {Finn}(1969)}]{Concus69}%
  \BibitemOpen
  \bibfield  {author} {\bibinfo {author} {\bibfnamefont {Paul}\ \bibnamefont
  {Concus}}\ and\ \bibinfo {author} {\bibfnamefont {Robert}\ \bibnamefont
  {Finn}},\ }\bibfield  {title} {\enquote {\bibinfo {title} {On the behavior of
  a capillary surface in a wedge},}\ }\href
  {http://www.pnas.org/content/63/2/292.abstract} {\bibfield  {journal}
  {\bibinfo  {journal} {Proceedings of the National Academy of Sciences}\
  }\textbf {\bibinfo {volume} {63}},\ \bibinfo {pages} {292--299} (\bibinfo
  {year} {1969})}\BibitemShut {NoStop}%
\bibitem [{\citenamefont {Snoeijer}\ and\ \citenamefont
  {Andreotti}(2013)}]{Andreotti2013}%
  \BibitemOpen
  \bibfield  {author} {\bibinfo {author} {\bibfnamefont {Jacco~H.}\
  \bibnamefont {Snoeijer}}\ and\ \bibinfo {author} {\bibfnamefont {Bruno}\
  \bibnamefont {Andreotti}},\ }\bibfield  {title} {\enquote {\bibinfo {title}
  {Moving contact lines: Scales, regimes, and dynamical transitions},}\ }\href
  {http://www.annualreviews.org/doi/abs/10.1146/annurev-fluid-011212-140734}
  {\bibfield  {journal} {\bibinfo  {journal} {Annual Review of Fluid
  Mechanics}\ }\textbf {\bibinfo {volume} {45}},\ \bibinfo {pages} {269--292}
  (\bibinfo {year} {2013})}\BibitemShut {NoStop}%
\bibitem [{\citenamefont {Gau}\ \emph {et~al.}(1999)\citenamefont {Gau},
  \citenamefont {Herminghaus}, \citenamefont {Lenz},\ and\ \citenamefont
  {Lipowsky}}]{Lipowsky99}%
  \BibitemOpen
  \bibfield  {author} {\bibinfo {author} {\bibfnamefont {H}~\bibnamefont
  {Gau}}, \bibinfo {author} {\bibfnamefont {S}~\bibnamefont {Herminghaus}},
  \bibinfo {author} {\bibfnamefont {P}~\bibnamefont {Lenz}}, \ and\ \bibinfo
  {author} {\bibfnamefont {R}~\bibnamefont {Lipowsky}},\ }\bibfield  {title}
  {\enquote {\bibinfo {title} {Liquid morphologies on structured surfaces: From
  microchannels to microchips},}\ }\href
  {http://science.sciencemag.org/content/283/5398/46} {\bibfield  {journal}
  {\bibinfo  {journal} {Science}\ }\textbf {\bibinfo {volume} {283}},\ \bibinfo
  {pages} {46--49} (\bibinfo {year} {1999})}\BibitemShut {NoStop}%
\bibitem [{\citenamefont {Speth}\ and\ \citenamefont
  {Lauga}(2009)}]{Lauga2009}%
  \BibitemOpen
  \bibfield  {author} {\bibinfo {author} {\bibfnamefont {Raymond~L}\
  \bibnamefont {Speth}}\ and\ \bibinfo {author} {\bibfnamefont {Eric}\
  \bibnamefont {Lauga}},\ }\bibfield  {title} {\enquote {\bibinfo {title}
  {Capillary instability on a hydrophilic stripe},}\ }\href
  {http://iopscience.iop.org/article/10.1088/1367-2630/11/7/075024} {\bibfield
  {journal} {\bibinfo  {journal} {New Journal of Physics}\ }\textbf {\bibinfo
  {volume} {11}},\ \bibinfo {pages} {075024} (\bibinfo {year}
  {2009})}\BibitemShut {NoStop}%
\bibitem [{\citenamefont {Tomotika}(1936)}]{Tomotika}%
  \BibitemOpen
  \bibfield  {author} {\bibinfo {author} {\bibfnamefont {S.}~\bibnamefont
  {Tomotika}},\ }\bibfield  {title} {\enquote {\bibinfo {title} {Breaking up of
  a drop of viscous liquid immersed in another viscous fluid which is extending
  at a uniform rate},}\ }\href {\doibase 10.1098/rspa.1936.0003} {\bibfield
  {journal} {\bibinfo  {journal} {Proceedings of the Royal Society of London A:
  Mathematical, Physical and Engineering Sciences}\ }\textbf {\bibinfo {volume}
  {153}},\ \bibinfo {pages} {302--318} (\bibinfo {year} {1936})}\BibitemShut
  {NoStop}%
\bibitem [{\citenamefont {Eggers}\ and\ \citenamefont
  {Villermaux}(2008)}]{Villermaux}%
  \BibitemOpen
  \bibfield  {author} {\bibinfo {author} {\bibfnamefont {Jens}\ \bibnamefont
  {Eggers}}\ and\ \bibinfo {author} {\bibfnamefont {Emmanuel}\ \bibnamefont
  {Villermaux}},\ }\bibfield  {title} {\enquote {\bibinfo {title} {Physics of
  liquid jets},}\ }\href {http://stacks.iop.org/0034-4885/71/i=3/a=036601}
  {\bibfield  {journal} {\bibinfo  {journal} {Reports on Progress in Physics}\
  }\textbf {\bibinfo {volume} {71}},\ \bibinfo {pages} {036601} (\bibinfo
  {year} {2008})}\BibitemShut {NoStop}%
\bibitem [{\citenamefont {Setu}\ \emph {et~al.}(2013)\citenamefont {Setu},
  \citenamefont {Zacharoudiou}, \citenamefont {Davies}, \citenamefont
  {Bartolo}, \citenamefont {Moulinet}, \citenamefont {Louis}, \citenamefont
  {Yeomans},\ and\ \citenamefont {Aarts}}]{Aarts2013}%
  \BibitemOpen
  \bibfield  {author} {\bibinfo {author} {\bibfnamefont {Siti~Aminah}\
  \bibnamefont {Setu}}, \bibinfo {author} {\bibfnamefont {Ioannis}\
  \bibnamefont {Zacharoudiou}}, \bibinfo {author} {\bibfnamefont {Gareth~J.}\
  \bibnamefont {Davies}}, \bibinfo {author} {\bibfnamefont {Denis}\
  \bibnamefont {Bartolo}}, \bibinfo {author} {\bibfnamefont {Sebastien}\
  \bibnamefont {Moulinet}}, \bibinfo {author} {\bibfnamefont {Ard~A.}\
  \bibnamefont {Louis}}, \bibinfo {author} {\bibfnamefont {Julia~M.}\
  \bibnamefont {Yeomans}}, \ and\ \bibinfo {author} {\bibfnamefont {Dirk G.
  A.~L.}\ \bibnamefont {Aarts}},\ }\bibfield  {title} {\enquote {\bibinfo
  {title} {Viscous fingering at ultralow interfacial tension},}\ }\href
  {http://pubs.rsc.org/en/content/articlelanding/2013/sm/c3sm51571k#!divAbstract}
  {\bibfield  {journal} {\bibinfo  {journal} {Soft Matter}\ }\textbf {\bibinfo
  {volume} {9}},\ \bibinfo {pages} {10599--10605} (\bibinfo {year}
  {2013})}\BibitemShut {NoStop}%
\end{thebibliography}

%


\end{document}